\documentclass[longauth]{aa}

\usepackage{graphicx}
\usepackage{txfonts}
\usepackage{hyperref}
\begin{document}

   \title{The Contribution of Alpha Particles to the Solar Wind Angular Momentum Flux in the Inner Heliosphere}

   \titlerunning{Alpha Particle Angular Momentum Flux}
   \authorrunning{Finley et al.}

    % \author[0000-0002-3020-9409]{Adam J. Finley}\exeter
    % \author[0000-0001-6077-4145]{Michael D. McManus}\ucb
    % \author[0000-0001-9590-2274]{Sean P. Matt}\exeter
    % % \author[0000-0002-2916-3837]{Victor R\'eville}\irap
    % % \author[0000-0001-8247-7168]{Rui F. Pinto}\irap
    % % \author[0000-0003-2061-2453]{Mathew Owens}\reading

    % \author[0000-0002-7077-930X]{Justin C. Kasper}\umich\sao
    % \author[0000-0001-6095-2490]{Kelly E. Korreck}\sao
    % \author[0000-0002-3520-4041]{A. W. Case}\sao
    % \author[0000-0002-7728-0085]{Michael L. Stevens}\sao
    % \author[0000-0002-7287-5098]{Phyllis Whittlesey}\ucb
    % \author[0000-0001-5030-6030]{Davin Larson}\ucb
    % \author[0000-0002-0396-0547]{Roberto Livi}\ucb

    % \author[0000-0003-1191-1558]{David M. Malaspina}

   \author{A. J. Finley\inst{1}
          \and
          M. D. McManus\inst{2}
          \and
          S. P. Matt\inst{1}
          \and
          J. C. Kasper\inst{3, 4}
          \and
          K. E. Korreck\inst{4}
          \and
          A. W. Case\inst{4}
          \and
          M. L. Stevens\inst{4}
          \and
          P. Whittlesey\inst{2}
          \and
          D. Larson\inst{2}
          \and
          R. Livi\inst{2}
          \and
          S. D. Bale\inst{2}
          \and
        %   J. W. Bonnell\inst{2}
        %   \and
          T. Dudok de Wit\inst{5}
          \and
          K. Goetz\inst{6}
          \and
          P. R. Harvey\inst{2}
          \and
          R. J. MacDowall\inst{7}
          \and
          D. M. Malaspina\inst{8,9}
          \and
          M. Pulupa\inst{2}
          }

    \institute{University of Exeter, Exeter, Devon, EX4 4QL, UK \\ \email{af472@exeter.ac.uk} \and
        University of California, Berkeley, CA, USA \\ \and
        University of Michigan, Ann Arbor, MI, USA \\ \and
        Smithsonian Astrophysical Observatory, Cambridge, MA, USA \\ \and
        CNRS and University of Orl\'eans, Orl\'eans, France \\ \and
        University of Minnesota, Minneapolis, MN 55455, USA \\ \and
        NASA/Goddard Space Flight Center, Greenbelt, MD, USA \\ \and
        Astrophysical and Planetary Sciences Department, University of Colorado, Boulder, CO, USA \\ \and
        Laboratory for Atmospheric and Space Physics, University of Colorado, Boulder, CO, USA}

   \date{Received August 30, 2020; accepted October 30, 2020}

% \abstract{}{}{}{}{}
% 5 {} token are mandatory

  \abstract{An accurate assessment of the Sun's angular momentum (AM) loss rate is an independent constraint for models that describe the rotation evolution of Sun-like stars.}{In-situ measurements of the solar wind taken by Parker Solar Probe (PSP), at radial distances of $\sim 28-55R_{\odot}$, are used to constrain the solar wind AM-loss rate. For the first time with PSP, this includes a measurement of the alpha particle contribution.}{The mechanical AM flux in the solar wind protons (core and beam), and alpha particles, is determined as well as the transport of AM through stresses in the interplanetary magnetic field. The solar wind AM flux is averaged over three hour increments, so that our findings more accurately represent the bulk flow.}{During the third and fourth perihelion passes of PSP, the alpha particles contain around a fifth of the mechanical AM flux in the solar wind (the rest is carried by the protons). The proton beam is found to contain $\sim 10-50\%$ of the proton AM flux. The sign of the alpha particle AM flux is observed to correlate with the proton core. The slow wind has a positive AM flux (removing AM from the Sun as expected), and the fast wind has a negative AM flux. As with previous works, the differential velocity between the alpha particles and the proton core tends to be aligned with the interplanetary magnetic field.}{In future, by utilising the trends in the alpha-proton differential velocity, it may be possible to estimate the alpha particle contribution when only measurements of the proton core are available. Based on the observations from this work, the alpha particles contribute an additional $10-20\%$ to estimates of the solar wind AM-loss rate which consider only the proton and magnetic field contributions. Additionally, the AM flux of the proton beam can be just as significant as the alpha particles, and so should not be neglected in future studies.}


   \maketitle
%
%-------------------------------------------------------------------

\section{Introduction}

During their main sequence lifetimes, Sun-like stars host magnetised stellar winds which steadily remove angular momentum (AM) causing their rotation periods to increase with age \citep[][]{lorenzo2019constraining, nascimento2020rotation}. This process is made more efficient by the presence of a large-scale magnetic field, which transfers AM (stored in magnetic field stresses) to the stellar wind particles (proton, alphas, etc) as the winds expand \citep{weber1967angular, mestel1968magnetic}. Therefore the AM-loss rates of these stars depend strongly on the stellar magnetic field strength/geometry \citep{matt2012magnetic, garraffo2015dependence, reville2015effect, pantolmos2017magnetic, finley2017dipquad, finley2018dipquadoct}, properties which are controlled by the stellar dynamo \citep[see review of][]{brun2017magnetism} whose activity is observed to depend strongly on stellar rotation \citep{wright2011stellar, wright2016solar}. A consequence of this activity-rotation relation is the observed convergence of rotation periods for Sun-like stars during the main sequence \citep[][]{gallet2013improved, gallet2015improved}, such that for stars older than around 1Gyr there is an approximate relationship between rotation period and stellar age \citep[e.g.][]{skumanich1972time}.

This relationship between rotation and age is a valuable tool in estimating stellar ages \cite[``gyrochronology'', e.g.][]{barnes2003rotational}, and (when combined with the activity-rotation relation) provides information about the past, present, and future circumstellar environment experienced by exoplanets orbiting a given star \citep{johnstone2015evolution, gallet2019rotational, johnstone2020active}. A similar manner of rotation period evolution is broadly observed for stars with masses less than $\sim 1.3M_{\odot}$ (low-mass stars), though the time taken to converge from an initial distribution of rotation periods varies \citep{matt2015mass, garraffo2018revolution, amard2019first}. As the rotation periods of Sun-like stars (and low-mass stars) can be estimated by monitoring their brightness variations (as starspots/faculae rotate into, and out of, view), their rotation-evolution has primarily been constrained by observing open clusters with known ages \citep{agueros2011factory, agueros2017setting, mcquillan2013measuring, nunez2015linking, rebull2016rotation, covey2016rapidly, douglas2017poking, curtis2019temporary}. However, the number of open clusters with ages between 2-10Gyr is currently insufficient to constrain the rotation period-age relationship \citep[see review of][]{bouvier2014angular}. Subsequently, asteroseismology (a technique capable of determining both the star's age and rotation rate) has been used to measure the rotation periods of 21 older main sequence stars (with ages from $1-10$Gyrs), the results from which imply a weakened AM-loss rate for stars at the age of, or older than, the Sun \citep[][]{van2016weakened, metcalfe2019understanding}. In addition to these rotation period observations, an estimate of the current solar AM-loss rate can be used as an independent constraint on the rotation-age relation for Sun-like stars near the solar age \citep[see discussion in][]{finley2018effect}.

%, which is critical for accurately evaluating the solar wind AM flux,

The solar wind AM-loss rate is difficult to constrain as the tangential speed of the solar wind at 1au is a few km/s \citep{nvemevcek2020non}, which is small compared to the average radial wind speed of 400-600km/s. Despite this, attempts to constrain the solar wind AM flux have been made in the past \citep[e.g.][]{lazarus1971observation, pizzo1983determination, marsch1984distribution, finley2019direct}. {Though each of these studies suffered from uncertainty either due to the precision of the spacecraft pointing or the ability of the plasma instruments to detect the small tangential flows. These uncertainties are lessened with \textit{Parker Solar Probe} (PSP) which samples the solar wind within 0.25au, where the signal to noise on the tangential wind speed measurements should be increased and the influence of wind-stream interactions reduced.} However during its first few encounters, PSP detected tangential solar wind speeds of up to $\sim 50$km/s \citep{kasper2019alfvenic}. These speeds are far greater than expected from magnetohydrodynamic (MHD) modelling \citep[e.g.][]{reville2020role} and, if they were prevalent throughout the corona, would be incompatible with the AM-loss rate implied by the slope of the rotation-age relation. \cite{finley2020solar} analysed the data from the first two orbits of PSP and showed these flows to be localised, existing along with wind streams that have large negative tangential speeds. Therefore, the average equatorial solar wind AM flux measured by PSP is similar to observations from previous spacecraft. At present, the mechanism(s) that generate these large tangential solar wind speeds have yet to be determined. One possibility is the motion/circulation of open magnetic flux at the base of the wind \citep[][]{crooker2010suprathermal, fisk2020global, macneil2020parker}, though this is also not well understood.

During PSP's first two perihelion passes (E01 and E02, respectively), neither the Solar Probe Cup \citep[SPC,][]{case2020solar} or the Solar Probe Analysers \citep[SPAN, Livi et al. submitted;][]{whittlesey2020solar} were able to confidently measure the properties of the alpha particles in the solar wind. Previously, \cite{pizzo1983determination} used \textit{Helios} observations to show that the alpha particles contain a large negative contribution to the total AM flux \citep[see also][]{marsch1984distribution}. Though, due to an error in spacecraft pointing, the data from \textit{Helios} required a significant correction, and more recent measurements of the alpha particle AM flux using the \textit{Wind} spacecraft at 1au instead suggest a much smaller (but still negative) contribution \citep{finley2019direct}. As the alphas particles carry roughly $15-40\%$ of the linear momentum in the solar wind, it is vital that their contribution to the total solar wind AM flux is better understood, in order to further constrain the slope of the rotation-age relation for Sun-like stars.

In this work, we analyse more recent SPAN-Ion observations from the third and fourth encounters of PSP (hereafter E03 and E04), which we use to determine the mechanical AM flux of both the solar wind protons and alphas particles. In addition, we use observations of the Interplanetary Magnetic Field (IMF) from the FIELDS instrument suite \citep{bale2016fields}, to evaluate the magnetic field stresses, so that the total solar wind AM flux can be determined. From the limited data available, we quantify the significance of the alpha particle AM flux in comparison to that of the protons and magnetic field stresses.

\section{Data}
\subsection{Fitting}
The SPAN-Ion instrument is an electrostatic analyser (ESA) that measures three-dimensional velocity distribution functions (VDFs) and is capable of distinguishing particle masses via time-of-flight measurements. {Three-dimensional counts spectra are organised into 32 energy-per-charge by 8 azimuthal angle by 8 polar angle bins. The 32 energy bins are logarithmically spaced from 125 eV to 20 keV, the 8 azimuthal angle bins {each} have a width of $11.25^\circ$, and the 8 polar angle bins {each} have an average width of $14.5^\circ$. It takes 0.218s to measure one complete 3D distribution. 16 proton VDFs and 32 alpha VDFs are summed together and stored in the L2 SF00 (proton) and SF01 (alpha) SPAN-Ion data products at cadences of 6.99s and 13.98s respectively.}

As of E04, a significant portion of the proton and alpha VDFs are still obscured by the spacecraft's heat shield and therefore lie outside SPAN’s field of view \citep[see Figure 8 of][]{kasper2016solar}. This leads to truncation of the VDF and significant inaccuracies in the y-components (in instrument coordinates) of the plasma moments. Since the instrument y-axis is the one most closely aligned with PSP’s tangential direction, this poses an obvious problem in calculating the AM flux. To this end, we performed bi-Maxwellian fits of the VDFs {using the Levenberg–Marquardt algorithm \citep{levenberg1944method,marquadt1963algorithm}}, in effect filling in the blocked parts of the VDF. {We parameterise a bi-Maxwellian distribution with the functional form,
\begin{equation}
f^M(\mathbf v) = n \left(\frac{m}{2\pi}\right)^{\frac{3}{2}} \sqrt{\frac{R}{T_\perp^3}} \exp \left( - \frac{\frac{1}{2}m(\mathbf{v} - \mathbf{\overline{v}})^2}{T_\perp} \left\{ \cos^2\Theta(R - 1) + 1\right\}\right),
\label{biMaxwell}
\end{equation}
where $n$ is the number density, $\mathbf{\overline{v}}$ the mean velocity, $T_\perp$ the perpendicular temperature, $R = T_\perp/T_\parallel$ the perpendicular to parallel temperature ratio, and $\Theta$ the angle between $\mathbf{v} - \mathbf{\overline{v}}$ and the magnetic field direction $\hat{\mathbf{B}}$, so that $\cos \Theta = ( \mathbf{v} - \mathbf{\overline{v}})\cdot \hat{\mathbf{B}} / |\mathbf{v} - \mathbf{\overline{v}}|$. For the protons we fit a bi-Maxwellian to both core and beam populations (denoted pc and pb respectively),
\begin{equation}
f_p(\mathbf{v}) = f^M_{pc}(\mathbf{v}) + f^M_{pb}(\mathbf{v}).
\label{protonfit}
\end{equation}}
{We constrain the proton beam component’s velocity to lie along the $\hat{\mathbf{B}}$ direction relative to the core, so that $\mathbf{v_{pb}} = \mathbf{v_{pc}} + v_d \hat{\mathbf{B}}$, where $v_d$ is the differential velocity between core and beam. There are thus 10 free parameters fit to the proton VDF. }

\subsection{Alpha Channel Contamination}
{The SF01 channel on SPAN-Ion should contain only alpha particles (as discriminated by time of flight measurements). However, a small proportion ($\sim 2-3\%$) of the protons were found to have leaked into this channel, appearing as spurious counts at lower energies, corresponding to the region of phase space occupied by the proton VDF in the SF00 channel (please see Appendix \ref{alphachannel} for some discussion of this issue). When fitting to this SF01 data product then, we include a term that corresponds to the contaminant protons, that is we fit a function of the form,
\begin{equation}
f_{\text{SF01}}(\mathbf{v}) = \epsilon f_p(\mathbf{v}) + f_{\alpha}^M(\mathbf{v}),
\end{equation}
where the first term corresponds to the spurious protons (and is the same expression as in equation (\ref{protonfit})), and the second term corresponds to a bi-Maxwellian fit (equation (\ref{biMaxwell})) to the true alpha particle component. When fitting to these SF01 counts spectra we do not fit again to the mean velocity or parallel and perpendicular temperatures of the proton core and beam components, rather we take the computed values of $\overline{v}, T_\perp$, and $T_\parallel$ from our fits to the proton beam and core, leaving them fixed, and only fit to the overall scaling (parameterized as $\epsilon$ in the expression above). It was checked that the contaminant protons did not have any energy and angle dependence, and so fitting to an overall scaling $\epsilon$ of the proton distribution was appropriate. $\epsilon$ thus represents the fraction of protons from the SF00 channel leaking into SF01 channel, which gives us a total of 7 free parameters to fit to (one for the effective contaminant density $\epsilon$, and 6 for the bi-Maxwellian describing the alpha particle distribution). An earlier method of measuring the alpha particle parameters made an attempt to first subtract the spurious protons before fitting (see Appendix \ref{alphachannel}). However we feel that the above method of fitting to both the contaminant protons and alpha particles at the same time, using fitted parameters from the SF00 channel, better accounts for overlap between the proton and alpha distributions, especially in slow solar wind. This is also important during the ubiquitous magnetic field reversals or switchbacks observed so far by PSP \citep{dudokdewit2019,mcmanus2020cross}, because the associated speed enhancements in the protons causes greater overlap of the proton and alpha VDFs in velocity space.
}

% and for the alphas we fit to a single bi-Maxwellian,
% \begin{equation}
% f_\alpha(\mathbf{v}) = f^M_{\alpha c}(\mathbf{v}).
% \label{alphafit}
% \end{equation}
% We constrain the proton beam component’s velocity to lie along the $\hat{\mathbf{B}}$ direction relative to the core, so that $\mathbf{v}_b = \mathbf{v_c} + v_d \hat{\mathbf{B}}$. No such constraint was placed on the alpha velocities relative to the protons. There are thus 10 free parameters fit to the proton VDF and 6 to the alpha VDF.

{Finally, spacecraft motion is removed from the fitted proton and alpha parameters to obtain plasma velocities in RTN coordinates. Fits were filtered for large residuals, or obviously unphysical fitted parameters, and discarded as necessary. Previous studies of the solar AM-loss have focused on the proton core population as it is thought to contain the majority of the proton AM flux, with the beam containing a small fraction of the proton mass and momentum (though this is not always the case, see Figure 3 of \citealp{stansby2020origin}). For this work we include the contribution of the beam in our calculation of the proton AM flux.}

\subsection{{Instrument Pointing Uncertainty}}
{The determination of the solar wind AM-loss rate is a very sensitive measurement, and, as pointed out by \cite{pizzo1983determination}, highly contingent on how well an instrument’s pointing direction is known. Writing a generic angular momentum particle flux term as $F_{AM} = r \rho v_r v_t = r\rho v_r^2 \tan \phi$, where $\rho$ is particle density, $r$ radial distance, $v_r,v_t$ the radial and tangential speeds respectively, and $\tan \phi = v_t/v_r$. Then the uncertainty in $F_{AM}$ due to an uncertainty in azimuthal angle $\phi$ is approximately given by $\delta F_{AM} = r\rho v_r^2\delta\phi$. An angular uncertainty $\delta \phi$ of $1^\circ$ is then seen to yield an uncertainty in $F_{AM}$ of the order of previous estimates of the solar wind AM flux. Any kind of systematic error in $\delta \phi$ will therefore pose a serious challenge to making a physically meaningful angular momentum measurement.}

{
The most restrictive constraints on instrument pointing directions on board PSP come from those required by the WISPR instrument \citep{vourlidas2016wide} to achieve its science objectives. WISPR’s pointing direction is required to be known to an accuracy of $0.1^\circ$ relative to its optical axis (John Wirzburger\footnote{PSP Spacecraft System Engineer}, Oct 2020, personal communication). In addition, the orientation of the spacecraft relative to the Sun centre is even more tightly constrained - it must be known to within $0.02^\circ$ relative to the WISPR optical z-axis. While such pointing direction constraints are not official science requirements for the SPAN-Ion instrument in the same way that they are for WISPR, the SPAN-Ion instrument team believes the pointing uncertainty to be well below $1^\circ$ (John Wirzburger, Oct 2020, personal communication). Even supposing SPAN-Ion pointing uncertainties were as much as 2-3 times larger than WISPR’s, this would still constitute an acceptable level of uncertainty allowing meaningful measurements of the solar wind AM flux to be made.
}

\subsection{Observations}
{The proton density measured by SPAN-Ion is underestimated in general compared to that measured by SPC, and needs to be scaled up to match. A required scaling up of the SPAN-Ion density is not unexpected because, of the plasma moments measured by ESA-type instruments, density is the most difficult to measure accurately since it depends most sensitively on instrumental parameters and calibration (this instrumental dependence almost entirely drops out of the expressions for higher order moments so this is not true for velocity, temperature, etc.). An instrument malfunction during E03 however meant that SPC densities were not available for most of this encounter, and E04 was the first time in the mission that the core of the proton VDF was more in SPAN-Ion’s field of view than SPC’s, thus making SPC’s density measurements in principle untrustworthy for most of the second half of E04 (as indicated by the quality flag). Because of this, and the fact that SPC densities are inter-calibrated with FIELDS quasi-thermal noise measurements, we use SPAN-Ion fitted densities of protons and alphas, scaled up by a factor of 1.25. The scaling factor should not differ between protons and alphas since the dependence on instrumental parameters is the same for both species.}
Additionally, the proton tangential speeds observed by SPAN-Ion and SPC are not always in agreement with one another, the root cause of which remains under investigation. For transparency, some comparison of the tangential wind speed in PSP's frame of reference is included within Appendix \ref{SPCvsSPAN}. As SPAN-Ion independently measures the tangential component of the flow, whereas SPC measures something closer to the flow angle in the plane of its detector plates, the velocities from SPAN-Ion are used as they are thought to be more reliable.

During E03, observations from SPAN-Ion are available from 27th August 2019 to 8th September 2019, with perihelion at $35.7R_{\odot}$ occurring on 1st September. Similarly for E04, observations cover 23rd January 2020 to 4th February 2020, with a perihelion at $27.9R_{\odot}$ (closer to the Sun by $7.8R_{\odot}$ than E03) on 29th January. When analysing quantities in this work, the data is averaged over three hour increments so that fluctuations, which are not representative of the bulk solar wind flow, and noise are reduced. Due to this averaging timescale, our choice of magnetic field data cadence from the FIELDS instrument suite has a negligible influence on our results, and so one minute cadence data is used. Any magnetic field measurements that were flagged by the instrument team as being bad/problematic are removed.

\begin{figure*}
   \begin{center}
    \includegraphics[trim=.cm .25cm .cm 0.cm, clip, width=\textwidth]{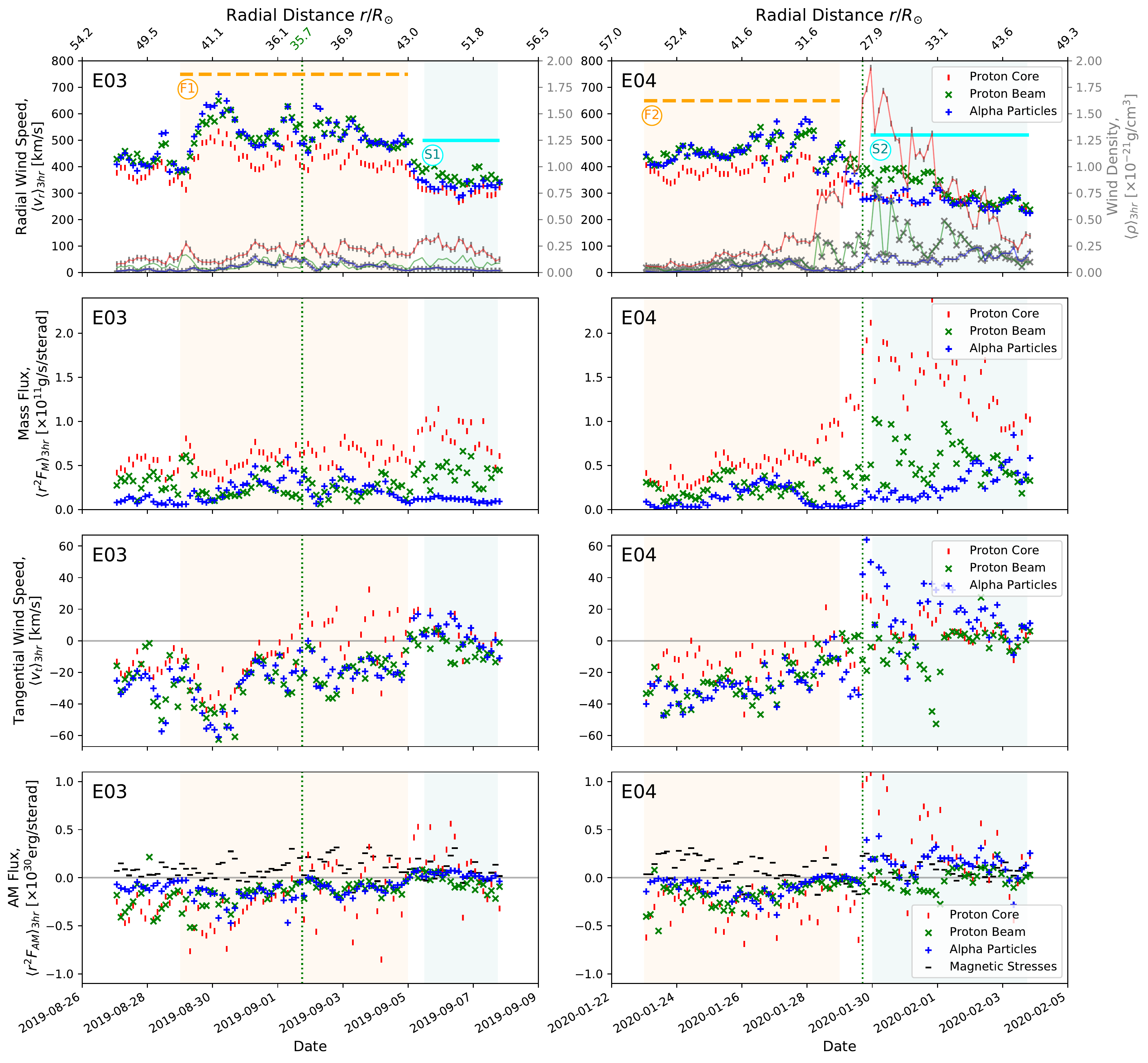}
  \end{center}
   \caption{Summary of the proton core (red vertical ticks), {proton beam (green crosses),} and alpha particle (blue plus signs) properties. {The first row displays their radial velocities and densities, which influence their mass fluxes in the second row. The particle mass fluxes, when combined with their respective tangential wind speeds (shown in the third row), produce the AM fluxes in the fourth row.} In the top panels, the particle densities have the corresponding symbol shapes but are grey in colour and connected by thin coloured lines. In the bottom panels, the magnetic stresses from equation (\ref{AMequation}) are shown with {black} horizontal ticks in comparison to the particle AM fluxes. The time of perihelion during each orbit is highlighted with a green dotted vertical line. Fast wind streams (F1 and F2) and slow wind streams (S1 and S2) are highlighted in orange and cyan, respectively.}
   \label{fluxes}
\end{figure*}

\begin{figure*}
   \begin{center}
    \includegraphics[trim=.cm .25cm .cm 0.cm, clip, width=\textwidth]{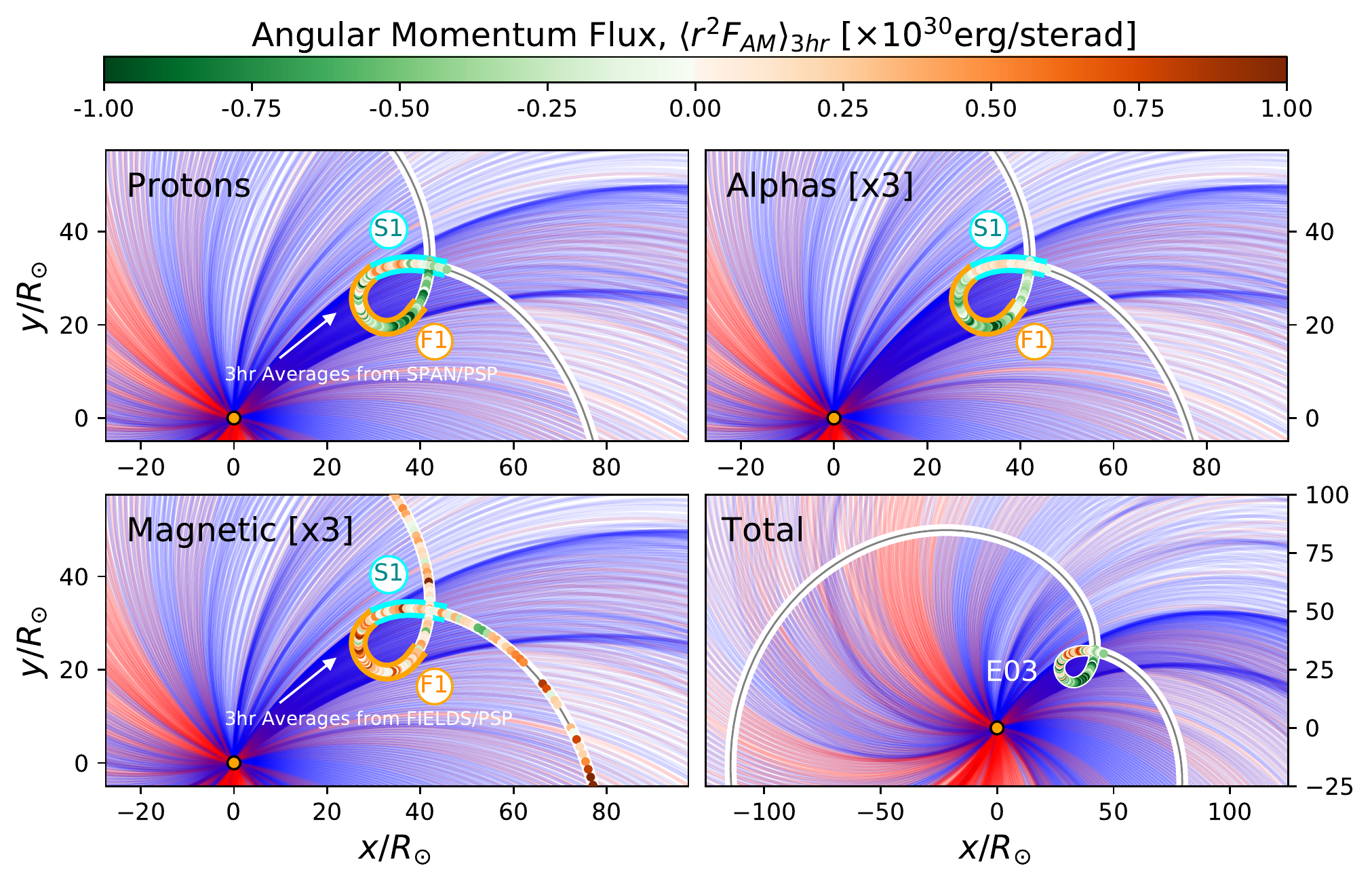}
  \end{center}
   \caption{The trajectory of PSP in a rotating frame of reference with the Sun, which when data are available, is coloured by the AM flux in the protons ({core plus beam,} top left), alpha particles (top right), magnetic stresses (bottom left), and their total (bottom right). Both the AM flux in the alpha particles and that stored in the magnetic stresses are multiplied by three, so that they are visible in comparison to the scale of the proton AM flux. In the background the average magnetic field polarity observed by FIELDS over one hour intervals colours parker spiral field lines, which are ballistically-mapped using the radial wind speed observed by SPC (as the data available from SPC covers a much longer time period than SPAN). A fast wind stream (F1) and a slow wind stream (S1) are highlighted in orange and cyan, respectively. The bottom right panel is zoomed-out slightly to show the extent of the observations in comparison to the system size. }
   \label{trajectoryE03}
\end{figure*}

\begin{figure*}
   \begin{center}
    \includegraphics[trim=.cm .25cm .cm .cm, clip, width=\textwidth]{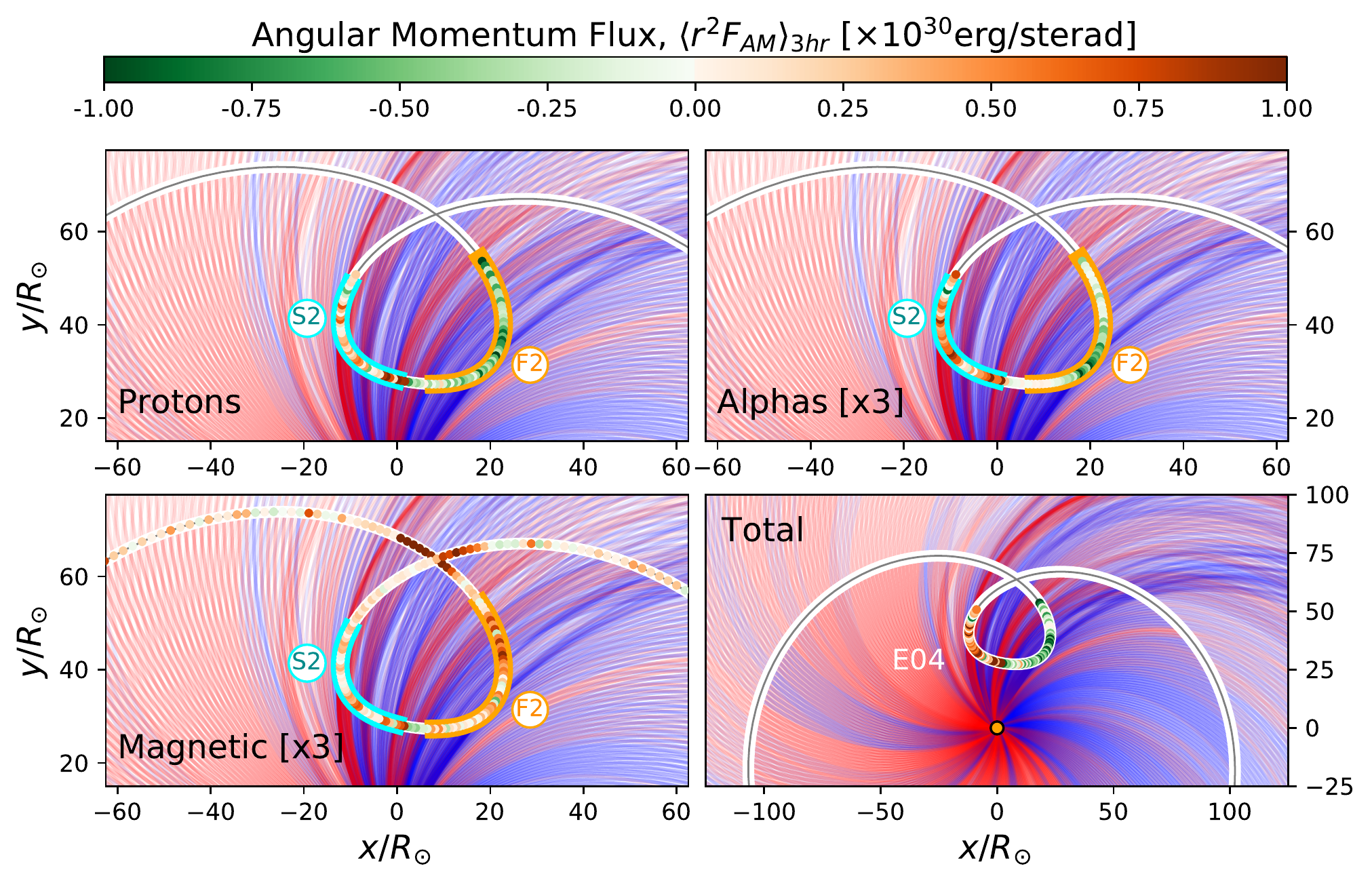}
  \end{center}
   \caption{Same as Figure \ref{trajectoryE03}, now with data from E04.}
   \label{trajectoryE04}
\end{figure*}

\begin{table*}
\caption{Average Properties of the Fast and Slow Wind Streams.}
\label{table:1}
\centering
\begin{tabular}{c c c | c c c | c }     % 7 columns
\hline\hline
Wind Stream & $\langle v_{r,\alpha}-v_{r,pc}\rangle$ & $\langle r^2 F_M\rangle$ & $\langle r^2F_{AM,p}\rangle$ & $\langle r^2F_{AM,\alpha}\rangle$ & $\langle r^2F_{AM,B}\rangle$ & $\langle r^2F_{AM}\rangle$\\
 & [km/s] & [$\times 10^{11}$g/s/sterad] & [$\times 10^{30}$erg/sterad] & [$\times 10^{30}$erg/sterad] & [$\times 10^{30}$erg/sterad] & [$\times 10^{30}$erg/sterad]\\
\hline
   F1 & 98 & 1.4 & -0.38 & -0.15 & 0.10 & -0.43\\
   S1 & 19 & 1.7 & 0.06 & 0.03 & 0.07 & 0.16\\
   F2 & 82 & 1.1 & -0.44 & -0.10 & 0.09 & -0.45\\
   S2 & -3 & 2.8 & 0.24 & 0.14 & 0.07 & 0.45\\  \hline
   $\langle$Fast$\rangle$ & 90 & 1.3 & -0.40 & -0.13 & 0.09 & $-0.44$\\
   $\langle$Slow$\rangle$ & 4 & 2.5 & 0.18 & 0.11 & 0.07 & $0.36$\\
\hline
\end{tabular}
\end{table*}

\section{Solar Wind Mass Flux and Angular Momentum Flux During E03 and E04}

As the solar wind particles remove AM at a rate proportional to their mass flux multiplied by their tangential speeds, it is informative to disentangle their relative abundance from the amount of AM they transport per unit mass. Throughout most of the E03 SPAN-Ion observations, PSP was immersed in relatively fast and low density wind (until the last few days where the wind speed decreased). The radial wind speed of both the protons and alpha particles are shown in Figure \ref{fluxes}, along with their densities. The mass flux ($F_M$) of each particle species during this time are evaluated using,
\begin{equation}
    r^2 F_M = r^2 (\rho_{pc} v_{r,pc} + \rho_{pb} v_{r,pb}) + r^2 \rho_{\alpha} v_{r,\alpha},
    \label{massflux}
\end{equation}
which is the sum of mass flux in the proton, and alpha particles multiplied by $r^2$ (corresponding to the flow of mass per solid angle). Here $r$ is the radial distance of PSP from the Sun, and the subscripts $pc$, $pb$, and $\alpha$ distinguish the proton core, proton beam, and alpha particle properties of density $\rho$, and radial wind speed $v_r$. The mass flux in each term of equation (\ref{massflux}) is displayed in Figure \ref{fluxes}. During E03, the proton mass flux remained roughly constant at $\sim 1.1\times 10^{11}$g/s/steradian {(with the beam carrying on average $\sim 31\%$ of the proton mass flux)}, whilst the alpha particle mass flux varied from $0.08-0.4\times 10^{11}$g/s/steradian. In the E04 SPAN-Ion observations, PSP encountered more varied solar wind conditions than E03 (this is also reflected in the IMF polarity during E04). Initially PSP was immersed in fast wind similar to that of E03, but then transitioned into slower and denser wind as PSP neared perihelion. Accordingly, the proton mass flux, as evaluated from equation (\ref{massflux}), increased two-fold. The alpha particle mass flux during E04 has some interesting features, including a large decrease prior to perihelion.

Combining the observations from SPAN-Ion and FIELDS, the solar wind AM flux ($F_{AM}$) is evaluated as a sum of the mechanical AM carried by the protons ($F_{AM,p}$), and alpha particles ($F_{AM,\alpha}$), along with a term prescribing the transfer of AM through magnetic field stresses ($F_{AM,B}$). This is given by,
\begin{align}
r^2F_{AM}=& r^2F_{AM,p} + r^2F_{AM,\alpha} + r^2F_{AM,B},\nonumber\\
=& r^3\sin\theta(\rho_{pc} v_{r,pc} v_{t,pc} + \rho_{pb} v_{r,pb} v_{t,pb}) + r^3\sin\theta\rho_{\alpha} v_{r,\alpha} v_{t,\alpha} \nonumber \\ &-r^3\sin\theta\frac{B_t B_r}{4\pi},
\label{AMequation}
\end{align}
where $\theta$ is its colatitude of PSP, $v_t$ denotes the tangential wind speed of each particle, $B_r$ is the radial magnetic field strength, and $B_t$ is the tangential magnetic field strength. A factor of $r^2$ is included, similarly to the mass flux in equation (\ref{massflux}), in order to remove any dependence of the AM flux on radial distance (note this quantity is now the flow of AM per solid angle, however it continues to be referred to as an AM flux for simplicity). {The tangential wind speeds of each particle species are shown in Figure \ref{fluxes}.} The resulting solar wind AM flux carried by each particle species during both E03 and E04 are shown in the bottom panels of Figure \ref{fluxes}, and in the context of PSP's location (in the rotating frame of the Sun) in Figures \ref{trajectoryE03} and \ref{trajectoryE04}. Note that PSP's latitude varies by a few degrees leading up to perihelion (heliographic latitude of around $-4^{\circ}$), which is not shown in Figures \ref{trajectoryE03} and \ref{trajectoryE04}. The structure of the IMF is depicted in the background by parker spiral magnetic field lines spaced an hour apart along PSP's trajectory which are ballistically-mapped from PSP using the average radial wind speed observed by SPC, and coloured by the average polarity of $B_r$ from FIELDS. The lower right panel of Figure \ref{trajectoryE03} indicates that during E03, PSP was broadly connected to the same side (negative $B_r$) of the Heliospheric Current Sheet (HCS). Whereas in Figure \ref{trajectoryE04}, the polarity of the IMF is much more variable, suggesting multiple crossings of the HCS. This is reflected in the range of wind speeds PSP encountered during E04, and the fluctuating abundances of the protons and alpha particles.

\begin{figure}
   \begin{center}
    \includegraphics[trim=.cm .25cm .cm .25cm, clip, width=0.45\textwidth]{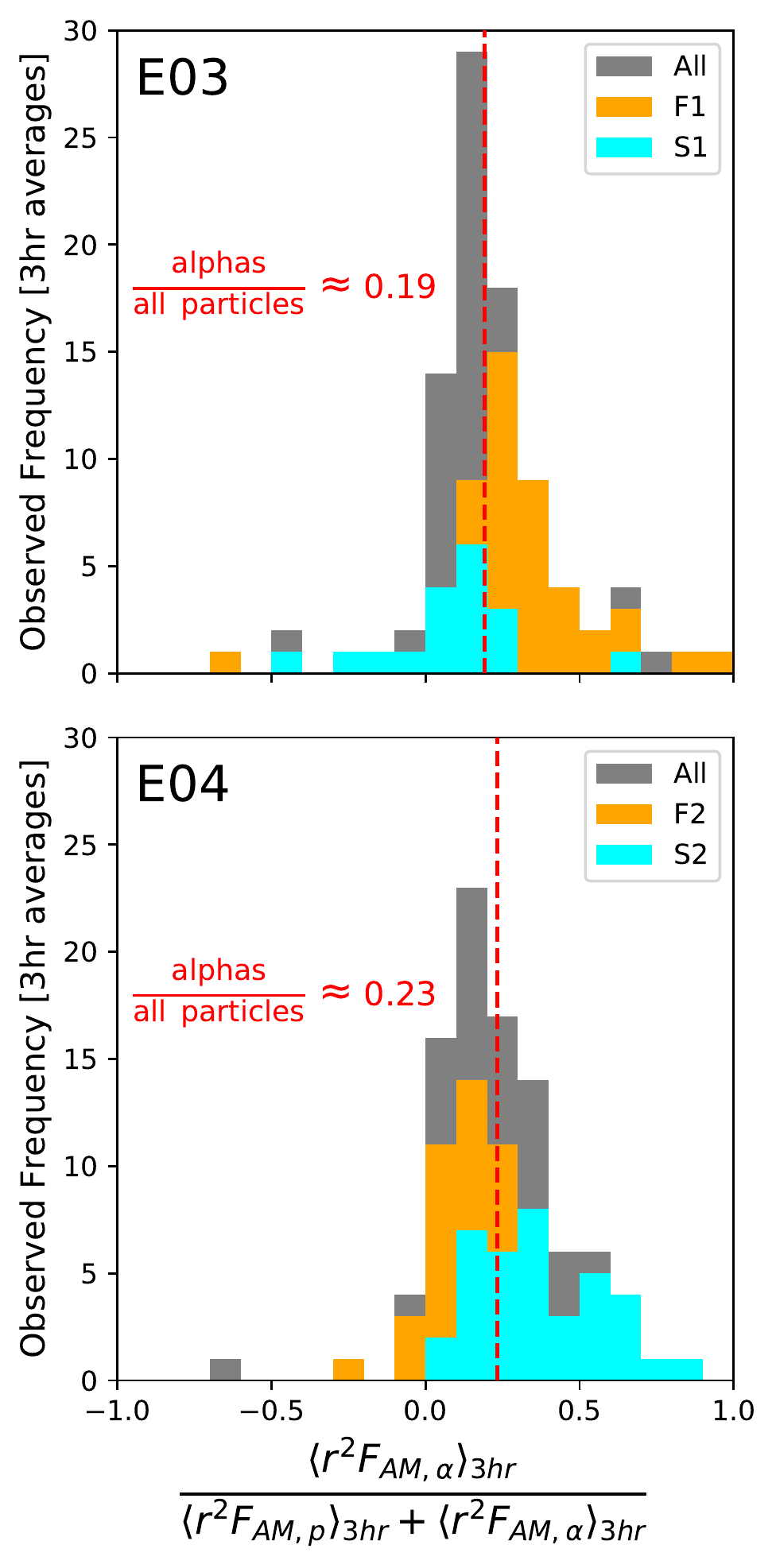}
  \end{center}
   \caption{Frequency of observing a given ratio of the alpha particle AM flux to the total mechanical AM flux {(proton core, proton beam, and alpha particles)} in the solar wind during E03 (top) and E04 (bottom). The average values from E03 and E04 are annotated and shown with red dashed lines. The alpha particles carry around a fifth of the mechanical AM flux in the solar wind. }
   \label{ratio}
\end{figure}

As previously noted in \cite{finley2020solar}, the magnetic stress term shows little variation compared to the proton AM flux. Figure \ref{fluxes} shows that, in general, the proton core carries the largest positive and negative AM fluxes, with the {proton beam and} alpha particle contributions being much less (in large part due to their lower abundance), and typically having the same sign as the proton core. {The proton beam has a similar behaviour in velocity to the alpha particles (and a similar magnitude of mass flux), therefore its contribution to the total AM flux is similar in size to that of the alpha particles. However, it is worth noting that the proton beam is the least well constrained component in the fitting process, and so throughout this work the total proton AM flux is favoured for comparison. The ratio of the alpha particle AM flux to the total mechanical AM flux, i.e. both the protons (core plus beam) and the alpha particles, is displayed as a histogram in Figure \ref{ratio}. Based on the observations from both E03 and E04, the alpha particles carry around $20\%$ of the mechanical AM in the solar wind (mean values of $19\%$ in E03, and $23\%$ in E04). Their contribution can vary significantly up to $50 - 80\%$, though there is a large number of observations with the alpha particles carrying less than $10\%$ of the total mechanical AM flux. }

\begin{figure*}
   \begin{center}
    \includegraphics[trim=.35cm .25cm .35cm .cm, clip, width=0.9\textwidth]{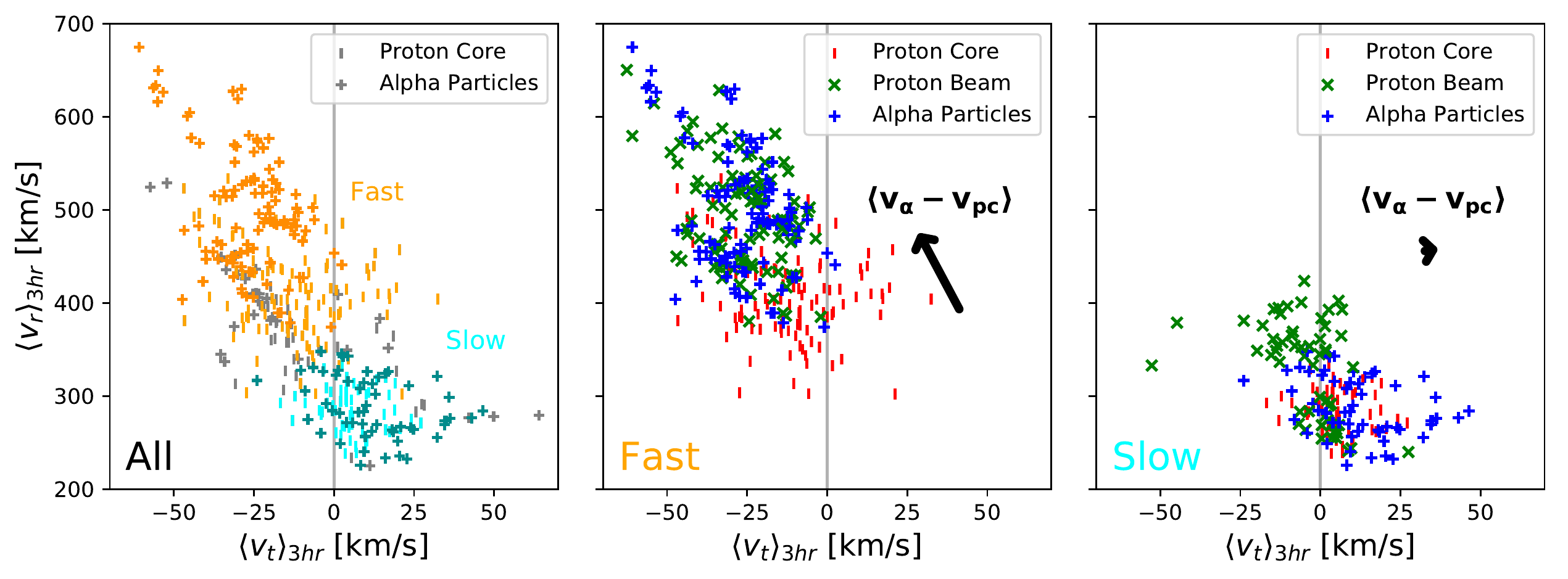}
  \end{center}
   \caption{Distribution of radial and tangential speeds for the solar wind protons and alpha particles. All the E03 and E04 observations of the proton core (vertical ticks), and alpha particles (plus signs) from SPAN-Ion are plotted in the left panel in grey, with the fast (F1 and F2) and slow (S1 and S2) wind streams highlighted. The distributions from the fast and slow wind streams are shown separately in the middle and right panels, respectively. {These include the proton beam (crosses) for completeness.} In each case, the average difference vector between the alpha particle and proton {core} velocities in the RT-plane are annotated {(shifted away from the data)}.}
   \label{velocityComp}
\end{figure*}

\section{Behaviour of the Alpha Particle Angular Momentum Flux}

\subsection{Fast Wind versus Slow Wind}
During the combined observations of E03 and E04, PSP was immersed in relatively fast wind ($v_r>450$km/s) more often than slow wind. During each encounter, one fast and one slow wind stream are identified (in orange and cyan, respectively) in order to further examine trends in the alpha particle AM flux. These time periods are highlighted, and numbered, in Figures \ref{fluxes}, \ref{trajectoryE03}, \ref{trajectoryE04} and \ref{ratio}. The averaged quantities from each stream are detailed in Table \ref{table:1}. {Figure \ref{ratio} shows that there is no consistent trend in the ratio of AM carried by the alpha particles (to the total mechanical AM flux) between the fast and slow streams.} On average in the two fast wind streams identified (F1 and F2), both the protons and the alpha particles are observed to carry a negative AM flux (i.e. adding AM to the Sun). {The distribution of the radial and tangential wind speeds of the protons and alpha particles are shown in detail in Figure \ref{velocityComp}.} The average tangential speeds in the fast wind are $\langle v_{t,pc}\rangle=-12$km/s, {$\langle v_{t,pb}\rangle=-26$km/s,} and $\langle v_{t,\alpha}\rangle=-26$km/s. In contrast to this, the slow wind streams (S1 and S2) appear to have a mostly positive AM flux in the protons and alpha particles (i.e. removing AM from the Sun, as expected). In this case their average tangential speeds are $\langle v_{t,pc}\rangle=5$km/s, {$\langle v_{t,pb}\rangle=-4$km/s,} and $\langle v_{t,\alpha}\rangle=12$km/s. In either wind conditions, the magnetic stresses are roughly the same and are a positive contribution to the total.

It is unlikely that the fast wind carries a net negative AM flux everywhere in the heliosphere, instead our observations are likely influenced by the large spatial and/or temporal variations in the equatorial AM flux \citep{finley2020solar}. Though a recent study of the solar wind at 1au by \cite{nvemevcek2020non} was able to show a similar correlation with the tangential speeds of the fast and slow wind \citep[see also Figure 2 of][]{finley2019direct}. Therefore, we cannot rule out the possibility that this dichotomy in AM fluxes (between fast and slow wind streams) exists more generally in the equatorial solar wind. Once more data has been collected by PSP, a statistical study of the AM flux carried by the fast and slow wind could determine the validity of this hypothesis. If this dichotomy were to exist, matching the large AM flux differences between the fast and slow wind would be a new observational constraint on models that aim to reproduce the solar wind acceleration and propagation \citep[e.g.][]{reville2020role, shoda2020alfv}. Currently, the source(s) of these enhanced tangential wind speeds remain unclear. As PSP moves closer to the Sun it becomes less likely that these features are a result of wind-stream interactions, and so perhaps a mechanism nearer the base of the wind is responsible, such as the motion of open magnetic flux in the photosphere \citep[e.g.][]{fisk2020global}.

\subsection{Alpha-Proton Differential Velocity}
The differential velocity between the alpha particles and the proton {core} (${\bf v_{\alpha} - v_{pc}}$), is observed to vary significantly between the fast and slow wind streams studied in this work. In the fast wind streams, the radial speed difference between the two particle species is around 90km/s, whereas for the slow wind streams its is nearer 5km/s. Previous observations of the solar wind at various distances from the Sun have shown that ${\bf v_{\alpha} - v_{pc}}$ is a function of wind speed, such that at low speeds the protons and alpha particles have similar velocities \citep{marsch1982solar, neugebauer1996ulysses}. At high speeds, ${\bf v_{\alpha} - v_{pc}}$ grows larger and becomes increasingly aligned with the direction of the IMF \citep{asbridge1976helium}. {Figure \ref{velocityComp} shows the sampled velocity space of the alpha particles and protons. Indeed the velocities of the alpha particles and proton {core} in the slow wind are far more similar than that of the fast wind. The average difference vector for the fast and slow streams are over-plotted in black.}

\begin{figure*}
   \begin{center}
    \includegraphics[trim=.3cm .25cm .5cm .cm, clip, width=\textwidth]{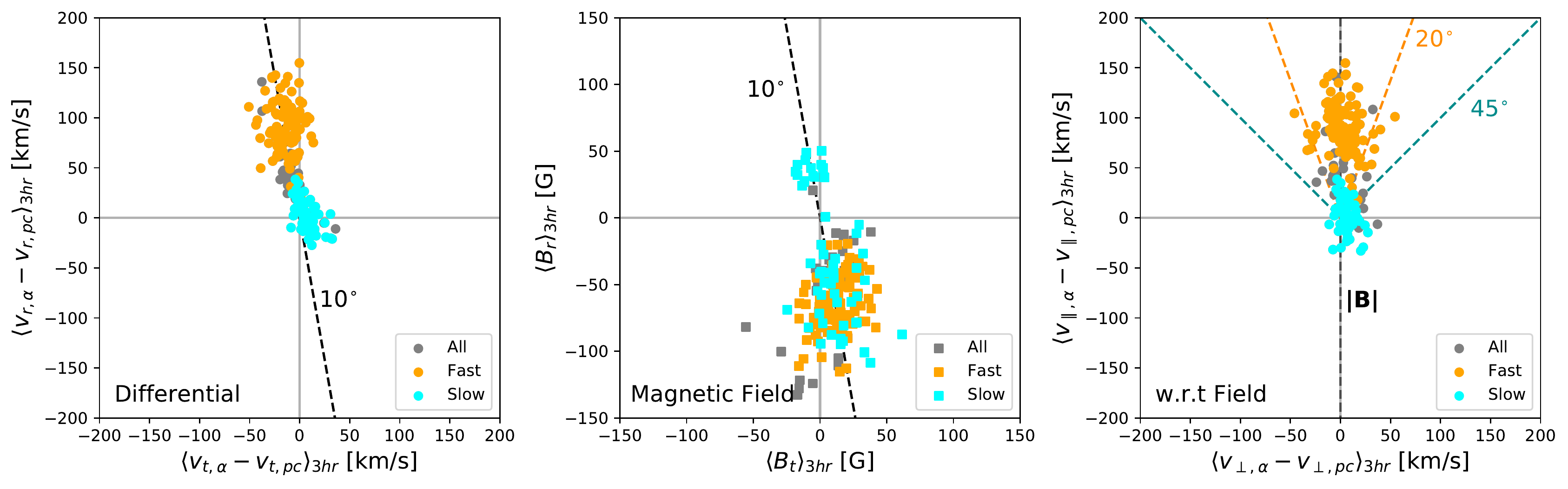}
  \end{center}
   \caption{Left: Distribution of the differential velocity vector between the alpha particles and proton {core} in the RT-plane. Middle: Distribution of IMF vectors in the RT-plane. Right: Distribution of the differential velocity vector between the alpha particles and proton {core} with respect to the IMF. The IMF points along the y-axis, the sign of the IMF (inward or outward from the Sun) is inconsequential. All observations are averaged over three hour intervals, as done previously when calculating the AM flux. Data from the fast (F1 and F2) and slow (S1 and S2) wind streams are highlighted in orange and cyan, respectively. }
   \label{DiffVelocity}
\end{figure*}

{Figure \ref{DiffVelocity} shows the direction of ${\bf v_{\alpha} - v_{pc}}$ in the RT-plane for each three hour average, along with the IMF direction (both of which are around $10^{\circ}$ from radial). The right panel of Figure \ref{DiffVelocity} shows the differential velocity vectors from the left panel with respect to the IMF direction from the middle panel. Given the relationship between the differential velocity and the IMF direction, it may be possible to infer the contribution of the alpha particles to the solar wind AM flux by rewriting $F_{AM,\alpha}$ as,}
\begin{equation}
    r^2 F_{AM,\alpha}= r^3\sin\theta \bigg(\cfrac{\rho_{\alpha}}{\rho_{pc}}\bigg)\rho_{pc}(v_{r,pc}+\delta v_{\alpha p, r})(v_{t,pc}+\delta v_{\alpha p, t}),
\end{equation}
{where $\delta v_{\alpha p, r}$ and $\delta v_{\alpha p, t}$ are the radial and tangential components of the differential velocity vector ${\bf v_{\alpha} - v_{pc}}$. Such that, given the abundance $\rho_{\alpha}/\rho_{pc}$ and a relation for ${\bf v_{\alpha} - v_{pc}}$ based on more readily observed parameters, e.g. ${\bf v_{\alpha} - v_{pc}} \approx f({\bf B},{\bf v_{pc}},...)$, the alpha particle AM flux could be estimated during times when they are not directly observed. However, as there are only a small number of observations available at present, it is difficult to say how reliable such a relationship for ${\bf v_{\alpha} - v_{pc}}$ would be. For F1 and F2, the angle between ${\bf v_{\alpha} - v_{pc}}$ is more strongly aligned with the IMF direction than during S1 and S2 (this can easily be seen in the right panel of Figure \ref{DiffVelocity}). Though the larger range of angles observed between ${\bf v_{\alpha} - v_{pc}}$ and the IMF during the slow wind streams is likely a consequence of small fluctuations between the particle speeds (including noise), as their velocity vectors are similar in size {due to the increased collisionality of the slow wind in comparison to the fast wind \citep[see][]{alterman2018comparison}}. Note that the proton beam generally has a similar velocity distribution in the R-T plane to that of the alpha particles\footnote{This is in-part because the differential velocity between the proton core and beam was previously constrained to lie along the magnetic field direction (see Section 2.1).}, and so when observations of the proton beam are available they too could be used to reconstruct the alpha particle AM flux. Though fits to the proton beam often carry higher uncertainties than for the proton core (due to the smaller number of counts in that region of phase space), and the proton beam population may not be present if the proton distribution is close to being isotropic.}

Consider the consequence of the differential velocity aligning with the IMF for a fast wind stream with a positive AM flux in the proton {core} (which is not observed here, but likely exists). For the differential velocity to be aligned with the IMF the alpha particles would have to take a smaller or negative tangential velocity. Though this has yet to be observed by PSP, it is capable of explaining previous observations from \textit{Helios} in which a positive AM flux in the proton core was observed with a negative AM flux in the alpha particles \citep{pizzo1983determination}.  Moreover, as the differential velocity is observed to lessen as the solar wind travels through interplanetary space \citep[see][]{neugebauer1996ulysses}, it is likely that AM is exchanged between the two species. So this could explain the observations from \textit{Wind} at 1au which show the alpha particles to have a much smaller contribution to the total AM flux \citep{finley2019direct}. However, the mechanism which produces the differential velocity between the alpha particles and the proton core is still unknown.

\section{Conclusion}

Using data collected by the SPAN-Ion and FIELDS instruments on-board PSP during its third and fourth closest approaches to the Sun, the strength of the AM flux carried by the alpha particles has been evaluated. Typically the alpha particles host around a fifth of the mechanical AM flux contained within the solar wind particles, the rest is carried by proton {core and beam}. Though it is difficult to say if this result is representative of the globally-averaged solar wind. The sign of the alpha particle flux follows closely that of the proton {core}. The fast wind is observed to carry a net negative AM flux, and the slow wind is found with a comparable but positive AM flux. However given the small number of available observations, it is difficult to say if this dichotomy is typical of the equatorial solar wind, or simply a reflection of the variability of the solar wind AM flux \citep[as discussed in][]{finley2020solar}.

For the data presented, there is a strong preference for the differential velocity between the alpha particles and proton {core} to be aligned with the IMF in the fast solar wind. For the slow wind, the alpha particles and protons are observed have more similar velocities. This was previously observed with \textit{IMP 6 $\&$ 7} \citep{asbridge1976helium}, \textit{Helios} \citep{marsch1982solar}, and \textit{Ulysses} \citep{neugebauer1996ulysses}. The evolution of the differential velocity with radial distance likely explains some of the differences between previous observations from \textit{Helios} \citep{pizzo1983determination}, and \textit{Wind} \citep{finley2019direct}. In future, trends in this quantity could be used to constrain estimates of the AM flux in the alpha particles when only the proton {core} is detected. Additionally, the proton beam velocity is shown to be similar to that of the alpha particles, so this too can be used to constrain the alpha particle AM flux (though measurements of the proton beam are typically more uncertain than the proton core). Interestingly, this work suggests that the proton beam contains a similar magnitude of AM flux to the alpha particles, and so (like the alpha particles) the beam should also be included in future estimations of the solar wind AM flux.

The average AM flux computed in this work would suggest that the Sun is gaining AM (i.e. spinning up). However this seems unrealistic, and is more likely due to: 1) PSP sampling only a small fraction of the equatorial wind, and 2) PSP primarily sampling the fast solar wind which is more often subject to dynamical interactions (i.e. deflections or compression/rarefactions {that generate a tangential flow in the solar wind) than the slow wind \citep[this can be seen in 1au measurements; e.g.,][]{finley2019direct, nvemevcek2020non}.}
Using a larger set of observations from PSP during its first two orbits (E01 and E02), the solar wind AM-loss rate (in just the protons and magnetic field stresses) was previously estimated to be $2.6\times10^{30}$erg and $4.2\times10^{30}$erg, respectively \citep{finley2020solar}. Assuming the observed fraction of the mechanical AM flux in the alpha particles is representative of the solar wind in general, then this previous estimate can be re-evaluated to include the alpha particle contribution {by increasing the proton AM flux by a quarter. As the ratio of the proton AM flux to magnetic stresses varied between orbits, this leads to an increase in the solar AM-loss rate of $12\%$ for E01, and $19\%$ for E02 (values of $2.9\times10^{30}$erg and $5.0\times10^{30}$erg, respectively).} So despite including the alpha particles, our estimate of the solar wind AM-loss rate remains smaller than the value of $6.2\times10^{30}$erg expected from a \cite{skumanich1972time}-like rotation-age relation for Sun-like stars \citep[see][]{finley2018effect}. Thus the slope of the rotation-age relation at the age of the Sun is currently steeper in rotation-evolution models than would be implied from observations of the solar wind.

With the solar cycle progressing towards increased solar activity, future works will be able to observe how the contribution of the alpha particles changes as their abundance increases \citep{alterman2020helium}. In addition, more frequent observation of the solar wind by both PSP, and \textit{Solar Orbiter} \citep[][]{mueller2013solar, muller2020solar, zouganelis2020solar}, will bring us closer to understanding the mechanisms that distribute the solar wind AM flux across the protons, alpha particles, and magnetic field stresses. \textit{Solar Orbiter} will also be able to sample the AM flux away from the solar equator, which will also help to constrain the solar AM-loss rate (as our current estimates rely on measurements of just the equatorial AM flux).

\begin{acknowledgements}
    {We thank the anonymous referee for their constructive comments which resulted in a more thorough discussion of the data products, the resulting velocity distributions, and the angular momentum flux carried by the proton beam population. These changes greatly increased the overall quality of this paper.}
    In addition, we thank the SWEAP and FIELDS instrument teams of Parker Solar Probe, and the NASA/GSFC's Space Physics Data Facility for providing this data.
    Parker Solar Probe was designed, built, and is now operated by the Johns Hopkins Applied Physics Laboratory as part of NASA’s Living with a Star (LWS) program (contract NNN06AA01C). Support from the LWS management and technical team has played a critical role in the success of the Parker Solar Probe mission.
    AJF and SPM acknowledge funding from the European Research Council (ERC) under the European Union’s Horizon 2020 research and innovation programme (grant agreement No 682393 AWESoMeStars).
    Figures in this work are produced using the python package matplotlib \citep{hunter2007matplotlib}.
\end{acknowledgements}


\bibliographystyle{aa}
\bibliography{papers}

\begin{appendix}
\section{Alpha Channel Contamination}\label{alphachannel}

\begin{figure*}
    \begin{center}
    \includegraphics[trim=.cm .cm .cm .cm, clip, width=1.0\textwidth]{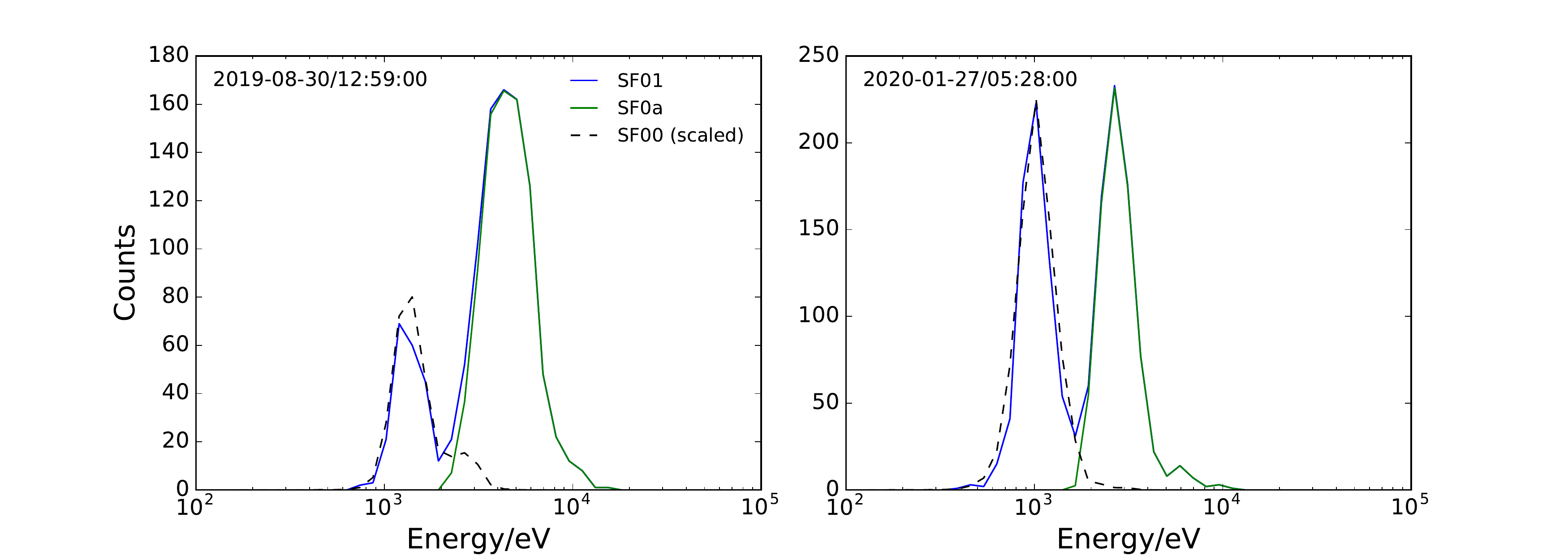}
    \end{center}
    \caption{Examples of counts versus energy per charge spectra during E03 (left plot) and E04 (right plot) from SPAN-Ion's SF01 (alpha particle) channel. Blue lines are the raw spectra, illustrating anomalous proton counts at low energies, dashed lines are SF00 proton counts scaled down to match the height of the lower energy peak, and green lines represent the corrected counts after applying our subtraction method.}
    \label{fig:alphachannel}
\end{figure*}

% Figure \ref{fig:alphachannel} shows two snapshots (one from E03 and one from E04) of counts vs energy per charge in SPAN-Ion’s SF01 channel, which should be the instrument’s alpha channel (as distinguished by time-of-flight measurements). The blue lines are the overall, uncorrected counts

As mentioned in the text, SPAN-Ion’s SF01 channel contains protons that have leaked in from the proton (SF00) channel. It is thought that these are due to so-called ``stragglers”, protons that aberrantly lose energy when passing through the ESA grids, thereby slowing down and getting counted as alphas. Figure \ref{fig:alphachannel} shows two snapshots, one from E03 and one from E04, of counts vs energy per charge (or equivalently, counts summed over the azimuthal and polar angles $\phi$ and $\theta$) in SPAN-Ion’s SF01 channel. The blue lines are the overall, uncorrected counts. The dashed line overlaid is the SF00 proton counts scaled down to match the height of the lower energy peak in each instance. This clearly shows that the low energy ($\sim 1$keV) counts in the SF01 spectra are due to protons, not alphas. There is nothing special about these times, a low energy ``shadow” of the proton distribution is always seen in the SF01 channel. An earlier version of this work involved first attempting to subtract the spurious protons, and then fitting the remaining alpha counts to extract alpha parameters. The method was as follows. We computed the ratio of counts between the SF01 channel and the SF00 channel, and averaged this ratio over a low energy region of phase space, (in particular, over the energy range from the lowest energy bin up to the energy bin one below the location of the primary proton peak), where the SF01 counts can be assumed to be almost all protons. This fraction of SF00 proton counts was then subtracted from the SF01 channel. We zeroed out counts at energies at or below the proton peak and, in the case of any energy bins which now contained negative counts after the subtraction, these were zeroed out as well. The green lines in figure \ref{fig:alphachannel} show the corrected counts (which we call ``SF0a") after applying this procedure, to which a bi-Maxwellian was then fit to. The later method, as explained in the text and used in the subsequent analysis, was to fit to both the proton and alpha parts of the SF01 spectra simultaneously, using our previously fitted proton core and beam velocities and temperatures as input, and to simply scale down the proton core and beam density as required to get the best overall fit.

% was simply to use the proton core and beam fitted velocities, temperatures, and densities to fit to the proton contaminant part of the SF01 spectra, then fit a bi-Maxwellian to the remaining (true) alpha counts, and simply scale down the proton core and beam density as required to get the best overall fit. This method is much simpler and utilises the fact that we have a clean proton channel from which we can tightly constrain the effective plasma parameters of the contaminating protons.

As can be seen from the two examples shown, the relative impact of the protons in SF01 varies considerably, depending on the abundance ratio $n_\alpha/n_p$, but also on the relative temperatures and drift speeds of the two species (in the E04 example, the protons can be seen to contribute a density as large as the density due to the alphas). In slow wind, where the differential speed between the two species is slower, a subtraction method can introduce larger artefacts in the alpha distributions due to the greater overlap in velocity space. It is worth noting however that our results did not significantly change at all between the two methods, because we are averaging over relatively long time intervals (3 hours).

% Very often the proton counts in this channel contribute a density that is larger than the density due to just the alpha particles, as can be seen in the example from E04.

% , showing the presence of contaminant protons at lower energies (around 1 keV). We compute the ratio of counts between the SF01 (alpha) channel and the SF00 (proton) channel, and average this ratio over a low energy region of phase space (in particular, over the energy range from the lowest energy bin up to the energy bin one below the location of the primary proton peak). This fraction of proton counts is then subtracted from the alpha channel. We zero out counts at energies at or below the proton peak and, in the case of any energy bins which now contain negative counts after the subtraction, these are zeroed out as well. The green lines in figure \ref{fig:alphachannel} show the corrected counts after applying this procedure, to which a bi-Maxwellian is then fit to. As emphasised in the text, this certainly introduces larger uncertainties in the lower energy bins (and barely changes the counts at the highest energies, as can be seen in the figure), but fitting mitigates this to a certain extent. The relative impact of the contaminant protons varies considerably, depending on the abundance ratio $n_\alpha/n_p$, but also on the relative temperatures and drift speeds of the two species. Very often the proton counts in this channel contribute a density that is larger than the density due to just the alpha particles, as can be seen in the example from E04.

\section{Comparison of the Tangential Velocities Measured by SPC and SPAN-Ion}\label{SPCvsSPAN}
\begin{figure*}
   \begin{center}
    \includegraphics[trim=.cm .cm .cm .cm, clip, width=.9\textwidth]{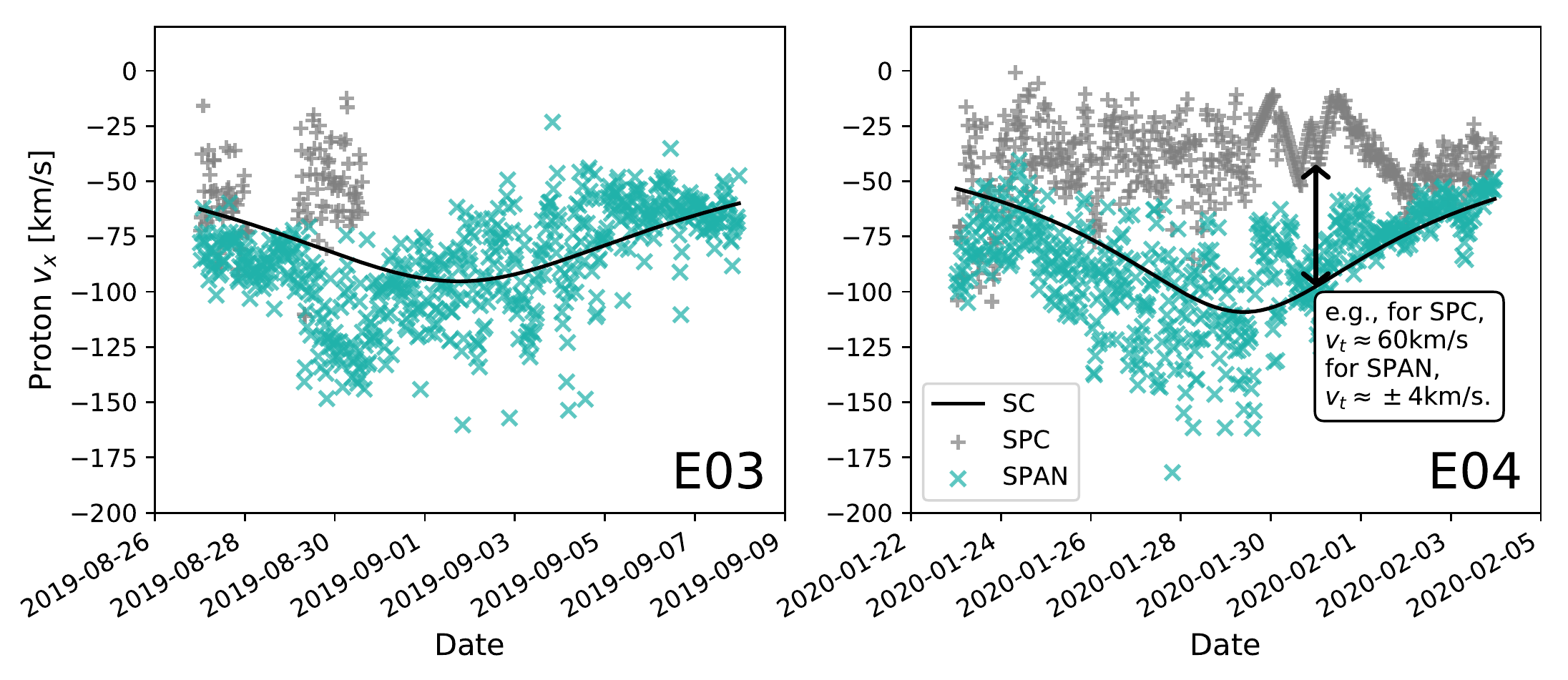}
  \end{center}
   \caption{Proton tangential speed as measured in the spacecraft frame ($v_x$) during E03 and E04 with SPC (grey `+') and SPAN (teal `x'). The tangential speed of PSP with respect to the flow (which must be subtracted from $v_x$ to recover $v_t$) is shown with a black line. The difference between the black line and the measured $v_x$ is the tangential speed $v_t$, as shown in the example.}
   \label{VXcomp}
\end{figure*}

The tangential speed of the protons in the solar wind was measured by both SPC and SPAN-Ion during E03 and E04. SPC suffered a technical failure part way through E03, and so only a few days are available for comparison. Additionally, given the substantial speed of PSP around the Sun in E04, the solar wind is not reliably measured by SPC for a period around closest approach. Figure \ref{VXcomp} displays 30 minute averages of the available observations in the spacecraft frame of reference (XYZ, where X=T, Y=-N, and Z=-R). The influence of the spacecraft's motion on this signal is shown by black lines in each panel. The tangential wind speed $v_t$ is obtained by subtracting the spacecraft's motion from these $v_x$ observations. As can be seen clearly during E04, SPC recovers a roughly constant $v_x$ between -10km/s and -50km/s, unlike SPAN-Ion which produces values much closer to the spacecraft motion, i.e. the tangential speed of the protons measured by SPAN-Ion is much smaller than that of SPC. As SPAN-Ion independently fits $v_x$, unlike SPC which measures something closer to the solar wind flow angle, we adopt values of $v_t$ from SPAN-Ion for the calculations in this work. Additionally using data from SPAN-Ion means that the proton and alpha particle velocities are subject to the same instrumental effects.

Despite the disagreement between SPC and SPAN-Ion during perihelion, there are times when the two instruments observe the same $v_x$, e.g. from 2nd February 2020 onward in E04. Additionally, variations in $v_x$ are recovered well by both instruments, e.g. 30th January - 1st February 2020. The cause for the overall discrepancy in $v_x$ is currently unknown, but as more data is accumulated by PSP a better understanding of both instruments will be achieved. It is also important to recall that the fits to SPAN-Ion data are performed on an incomplete VDF (see Section 2.1), therefore it is not currently possible to determine which instrument is measuring the flow correctly.  At present, this comparison can only be performed with data from E04, and so it is difficult to say how this disparity influences previous results from E01-E03 (which also have a different orbit to E04). Though the limited data from E03 suggests a similar departure between the two instruments on-board PSP.

% \section{Average Alpha-Proton Differential Velocity Vectors}\label{schematic}

% Figure \ref{schematic} shows the relationship between the differential velocity $\bf \langle v_{\alpha}-v_p\rangle$ and the IMF vector $\bf B$, for average values in the fast and slow wind streams. The dichotomy between the fast and slow wind shown in this work will likely not persist as PSP gathers more observations. For example, during E02 a fast wind stream was observed during perihelion that had a positive AM flux in the protons (see Finley et al. Submitted). In this case we would expect the alpha particles to travel at a velocity which aligns $\bf \langle v_{\alpha}-v_p\rangle$ with the IMF, and thus their tangential speeds may have been small or negative.

% \begin{figure*}
%   \begin{center}
%     \includegraphics[trim=.cm 3.cm .cm 3.cm, clip, width=0.9\textwidth]{f6.pdf}
%   \end{center}
%   \caption{Schematic representation of the average fast (left) and slow (right) wind stream vectors. In RTN coordinates, the angle of the IMF is shown in the background with thin grey lines. Over-plotted are the average proton $\langle v_p\rangle$, and alpha particle $\langle v_{\alpha}\rangle$, velocity vectors (confined to the RT-plane) in red and blue respectively. The alpha-proton differential velocity vectors $\langle v_{\alpha}-v_p\rangle$ are shown to scale with coloured arrows. }
%   \label{schematic}
% \end{figure*}
\end{appendix}

\end{document}